\documentstyle[aps,twocolumn,psfig]{revtex}   

\begin{document}
\title{Mesoscopic description of the annealed Ising model, and 
multiplicative noise}
\author{Walter Genovese$^{1}$, Miguel A. Mu{\~{n}}oz $^{1,2}$,
and P. L. Garrido $^{2}$
}
\address{ $^{1}$ Dipartimento di Fisica,
Universit\'a di Roma ``La Sapienza'',~P.le
A. Moro 2, I-00185 Roma, Italy}
\address{ $^2$ Instituto Carlos I de F{\'\i}sica Te{\'o}rica y Computacional,
Universidad de Granada, E-18071 Granada, Spain.}
\date{\today }
\maketitle

\begin{abstract}
A new type of Langevin equation exhibiting a non trivial
phase transition associated with the presence of multiplicative
noise is introduced. The equation is derived as a mesoscopic
representation of the microscopic annealed Ising model (AIM)
proposed by Thorpe and Beeman, and reproduces perfectly
 its basic phenomenology.  
The AIM exhibits a non-trivial behavior as the temperature is increased,
in particular it presents a disorder-to-order phase transition at low
temperatures, and a order-to-disorder transition at higher temperatures.
This behavior resembles that of some Langevin equations
with multiplicative noise,
 which exhibit also two analogous phase transitions
as the noise-amplitude is increased.
By comparing the standard models for noise-induced transitions with our 
new Langevin equation we elucidate that the mechanisms controlling the 
disorder-to-order transitions in both of them are essentially different,
even though for both of them the presence of multiplicative noise is     
a key ingredient.
\vspace{8pt}
\vspace{8pt}

\end{abstract}

\begin{text}
\narrowtext

\section{Introduction}

   A great deal of attention has been recently devoted to the study
of physical effects induced by the presence of noise, i.e. phenomena
appearing in stochastic systems, which would be absent in the sole
presence of the deterministic part of the corresponding Langevin
 equation \cite{HL}.
 By now it is clear that noise can generate quite unexpected 
and counterintuitive behaviors as, for example, the  {\it stochastic 
resonance} \cite{sr}, in which the output to input ratio  of a 
bistable system subjected to the presence of an oscillating force
 is strongly
enhanced by the presence of an additional stochastic term. Other 
examples are the resonant activation \cite{sa}, and
the noise induced spatial patterns \cite{n1}.
Another type of phenomena the noise is at the base of, are the so called
{\it  noise induced phase  transitions}. These came to light in
 an interesting paper by Van der Broek, Parrondo and Toral 
\cite{Raul} (see
also  \cite{HL,BK,Pik}).
These authors pointed out the fact that some Langevin equations 
may exhibit a noise-induced ordering transition (NIOT), i.e. a phase 
transition that is not expected from the analysis of the deterministic
part of such equation.  The phenomenology is as follows:

 i) For low 
enough noise amplitudes the system is disordered
 (i.e. the order parameter
takes a zero value).

 ii) At a certain critical value of the noise 
amplitude the system exhibits a NIOT and, in a range of
noise intensities above it the system remains ordered.

 iii) Finally, for noise
amplitudes larger than a second critical value, the noise operates 
in a more standard way, this is, disordering again the system. We
refer to this second phase transition as noise induced disordering
transition (NIDT).

A physical explanation
 of the NIOT  was 
proposed in \cite{Raul}; The ordering of the system is the consequence
of the interplay between the noise and
 the spatial coupling \cite{Kawai}. In particular, the noise 
generates a short time instability at every single site, and 
the presence of a spatial coupling renders stable the non trivial 
state generated in that way.

   A minimal model capturing the essence of the NIOT has been 
recently proposed \cite{GMs}. It has been clarified that the 
essence of the NIOT is purely multiplicative, this is, 
in order to generate 
an ordering transition the noise has to appear multiplied 
by the field variable. In this way, it has been possible to 
recognize
that the NIOT is characterized by a set of critical exponents
other than those 
of well established universality classes (as for example,
that of the Ising model) \cite{Noi,GMs}. 
 It has also been shown that 
due to the multiplicative origin of this transition it is possible 
to observe the phenomena in $d=1$, dimension at which is very  
unusual to observe phase transitions.

   Other results concerning NIOTs  and NIDTs can be found in the
literature \cite{Sancho,Muller,Kim}. A common feature of all the 
previously referred models, is that they are defined by means of
Langevin equations,
this is, equations describing the physics 
at a mesoscopic, coarse-grained scale (in fact, the concept of noise
is meaningful only at this level). 
In this context, it is an interesting task that of analyzing microscopic
models that exhibit similar non-trivial behaviors; one of which is
the anneal Ising model \cite{TB}.
By studying the connection between 
a microscopic systems and their respective mesoscopic representation 
one could shed some light 
on the way in which microscopic mechanisms generate the
very non-trivial 
effects described at a mesoscopic scale.

 In what follows we introduce the time honored anneal Ising model. 
It was proposed and described more than twenty years ago 
by Thorpe and Beeman \cite{TB}. A more detailed description of it
 will be presented in the 
next section; here we summarize the main
 properties we are interested in.
 The system is an Ising model in which the interactions, $J$, among spins,
are annealed (not quenched) random variables 
that change from bond to bond and are
 extracted from a fixed probability distribution, $P(J)$.
Under certain conditions (this is, for some distributions $P(J)$ to 
be specified later),
the system  
phenomenology is as follows:

 i) For low temperatures the system is disordered,
i.e. the averaged magnetization is zero. 

ii) At a critical value of the temperature, $T_1$,
the system exhibits a second order phase transition.             
 As the temperature is further increased above $T_1$ 
the averaged magnetization keeps on growing until it reaches 
a  maximum value and it starts decreasing if $T$ is increased further. 

iii) At a second critical temperature, $T_2$, the system exhibits
 another phase transition
 (analogous to the well known ferromagnetic-paramagnetic
disordering transition of the standard pure Ising model). 
The system remains disordered
for temperatures higher than $T_2$.

   This phase diagram resembles very much the         
 behavior of the previously described noise induced transitions
in Langevin equations.
 It is our purpose here to analytically
derive a coarse-grained, mesoscopic,
representation, in terms of a Langevin equation, 
of the microscopic annealed
Ising model (AIM) to further explore the eventual relations between both
phenomena.

\section{The annealed Ising model}

  Let us consider a d-dimensional impure Ising model in the sense that 
the value of the coupling constant among spins, $J$,
 changes from bond to bond, being an annealed random variable
with a fixed temperature-independent
probability distribution, $P(J)$ (which is not quenched but
annealed at every site).
 Following the strategy proposed by Thorpe and Beeman  \cite{TB}
the model can be exactly mapped into a standard
pure Ising model with  an effective 
parameter, $K =J/T$, that depends on $P(J)$ and $T$, and we write as 
$K_{eff}(T)$.
 In particular \cite{TB},
\begin{equation}
\int dJ \frac{P(J)}{coth[ K_{eff}-  J/T ]-\epsilon
 (K_{eff})}=0
\label{eff}
\end{equation}
where $\epsilon(K)$
  is the correlation function of two nearest-neighbor
spins in the pure Ising model. By solving the implicit equation 
Eq. (\ref{eff}) one obtains $K_{eff}$ as a function of the 
 temperature and the parameters characterizing $P(J)$. Note 
that, in particular, for the two-dimensional case, the Onsager's 
solution \cite{Ons} provides an explicit
 value for $\epsilon(K)$ and therefore
 Eq. (\ref{eff}) can be solved and, furthermore, the
system magnetization can be expressed as a function of $T$.
  Let us suppose now that, in particular, the distribution $P(J)$
is centered at a positive value $J_0$ (favoring ferromagnetic ordering),
and has a variable width (standard deviation), $\delta J$.
The resulting magnetization 
for this particular type of distribution is qualitatively 
represented in figure 1 (see also \cite{TB}). 

\begin{figure}
\centerline{\psfig{figure=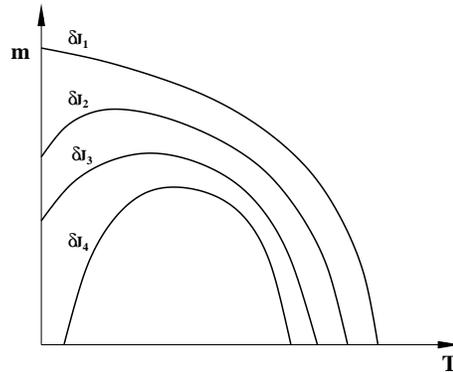,width=6cm}}
\caption{Magnetization as a function of the temperature $T$, for
different values of $\delta J$; $0 = \delta J_1 < \delta J_2 <
\delta J_3 < \delta J_4$ for the annealed Ising model. 
}
\end{figure}
Observe that for narrow distributions of $J$ the magnetization curve is
similar to its corresponding in the pure Ising model. Instead, 
as $\delta J$ is increased, a disordering tendency is observed at low   
temperatures, and in particular, for a large values of the width (as 
for example, $\delta J_4$ in Fig.1)
 the system is disordered at low temperatures, 
and exhibits a disorder-to-order phase transition
 at a certain temperature. 
The standard
ferromagnetic-paramagnetic (order-to-disorder) transition is also 
present and occurs at a variable value of $T$ for different values
of $\delta J$.

  The physical mechanism leading to the previous behavior was argued
in \cite{TB} to be the competition between ferromagnetic
 and antiferromagnetic
types of interactions that emerges
when sufficiently large values of $\delta J$ are considered.
In particular, when $\delta J> J_0$
both positive and negative values of the coupling constant are accessible
at each bond, and in that case , for low temperatures the system 
is in a {\it frustrated state} in which ferromagnetic and 
antiferromagnetic domains compete.
That frustration makes  
the ferromagnetic order parameter to vanish. 
As the temperature is further raised the thermal
noise activates annihilation of domain walls and the system is more
likely to ordinate. As a result, the averaged magnetization grows
with increasing temperature. At a given point this effect ceases,
and the standard role of the temperature as 
a disorganizing source sets to work.
 
 \section{Continuous representation}

    Let us now follow a standard procedure \cite{Amit}
 to cast the previous 
AIM into a continuous Langevin equation. For that purpose we first consider
the pure Ising model case, and write
down its associated equilibrium partition function:  
\begin{equation}
Z=\sum_{\{s\}}exp\left( \sum_{ij}K_{ij}s_{i}s_{j}\right).
\label{amit1}
\end{equation}      
Introducing auxiliary Gaussian integrals in terms of
 continuous variables
 $\phi_i$ (with $i$ varying from 1 to the total number of
 spins, $N$, in the lattice), and
performing the change of variables $\psi_i=  K_{ij}^{-1}\phi_j$
 we obtain \cite{Amit}
\begin{equation}
Z\propto \int d{\psi}_{1}...d{\psi}_{N}exp\left[-\frac{1}{4}{\psi}_{i}
K_{ij}{\psi}_{j}+\sum_{i} \log \cosh(K_{ij}{\psi}_{j})\right].
\label{amit2}
\end{equation}
 Expanding the hyperbolic-cosine in power series, performing a 
transformation to
Fourier space,  considering only the leading dependence on the temperature,
and performing the continuous limit 
we finally obtain  \cite{Amit}
 \begin{eqnarray}
Z & \propto &\int d[\psi] \ e^{-H} \nonumber \\
H &=&  \frac{1}{4} \int d^{d}x \
 [K_{0}(1-2K_{0}){\psi}^{2}(x)+{\rho}(4K_{0}-1){(
\nabla \psi)}^{2}  \nonumber \\
 &+&  \frac{1}{3}{K_{0}}^{4}
{\psi}^{4}(x)].
\label{amit2b}
\end{eqnarray}
with $K_0 = \int d^d x  K(x)$, and $\rho=1/2 \int d^d x K(x) x^2 $.
In this way we have derived a Ginzsburg-Landau coarse grained Hamiltonian
for the Ising model. This could have been guessed a priori by using 
heuristic arguments, but we have preferred to follow the previous
 procedure that permits
to obtain explicit expressions for the coefficients as function of the
microscopic parameters.
 In this way, observe, for example,
that both the diffusion constant and the coefficient of the quadratic
term depend on the coupling through $K_0$, therefore in order to 
simplify the notation we define the diffusion constant, $D=\rho (4 K_0 -1)$.
Taking only the main relevant dependences on $D$ we can write
\begin{equation}
H=\int d^{d}x \
 \left[\frac{a D}{2}{\psi}^{2}+\frac{b}{4}{\psi}^{4}+\frac{D
}{2}({\nabla \psi})^{2}\right],
\label{amit3}
\end{equation}
where $a$ is a tuning parameter proportional to the distance 
to the critical temperature,
and $b$ is a positive parameter.
Let us stress once more that we are neglecting higher order dependences
of $b$ and $a$ on $D$, and we assume them to be unessential to reproduce 
the microscopic phenomenology of interest at mesoscopic level 
(this hypothesis will be verified afterwards).
The simplest Langevin equation with a
stationary distribution characterized by a Gibbsian distribution with 
the Hamiltonian in Eq. (\ref{amit3}) is well known to be 
\cite{HH,Gardiner}
 \begin{equation}
{\partial}_{t} \psi =-(aD+b{\psi}^{2})\psi +D{\nabla}^{2}\psi+\eta(t)
\label{amit4}
\end{equation}
where $\eta(t)$ is a Gaussian white noise with $\langle \eta(x,t) \rangle = 0$,
and
$\langle \eta(x,t) \eta (x',t') \rangle =  \delta^d(x-x') \delta(t-t')$.

  At this point we can analyze the effects of an annealed distribution of $J$ 
in the microscopic AIM 
at the level of Langevin equations. For that purpose let us
 observe that in order to mimic the variability of the coupling in the
AIM we can just
substitute $D$ at each site
 in Eq. (\ref{amit4}) by a stochastic variable, namely:
$ D \rightarrow D+{\xi(x,t)} $, with $\langle \xi(x,t) \rangle = 0$ and
$\langle \xi(x,t) \xi (x',t') \rangle = \sigma_D^2 \delta^d(x-x')
 \delta (t-t')$, where $D$ and ${\sigma}_{D}$ play the role of $J_0$ and
$\delta J$ respectively in the microscopic model. 
In this way we obtain,
\begin{equation}
{\partial}_{t} \psi 
=-[a(D+{\xi})+b{\psi}^{2}]\psi +D{\nabla}^{2}\psi+\nabla
 ({\xi}\nabla \psi)+\eta(t).
\label{final}
\end{equation}
This equation (intended in the Ito interpretation \cite{Gardiner})
constitutes our continuous representation of the AIM.
Let us underline that the differences with respect to the pure case,
Eq. (\ref{amit4}, are two: the presence of a {\it multiplicative noise},
and an extra term that couples spatial fluctuations of $D$
with $\nabla \psi$.
Changes of $a$, parameter which appears multiplying
both the linear term and the multiplicative noise,
correspond to temperature variations.

   We have analyzed  Eq. (\ref{final}) in mean field approximation 
\cite{Kawai,Max,GMs}, and by performing systematic numerical simulations in
two dimensions. The mean field approximation is performed along the
lines discussed in \cite{Kawai,Max,GMs}. For the numerical simulation
we have employed the Euler method \cite{Max},
in a 32*32 lattice, with lattice spacing $\Delta a=1$, and considered 
a time mesh $\Delta t=0.001$.  Without lost of generality 
the parameters $b$ and $D$ have been fixed 
to $1$ and $10$ respectively.  Different noise amplitudes, $\sigma_D$,
 have been considered.
The main results we have obtained are as follows:
in both, mean field approximation and in the numerical simulation,
we reproduce the qualitative behavior of the order parameter as
a function of the temperature characteristic of the microscopic model
(see Fig.2 and Fig.3 and compare them with
Fig.1).

\begin{figure}
\centerline{\psfig{figure=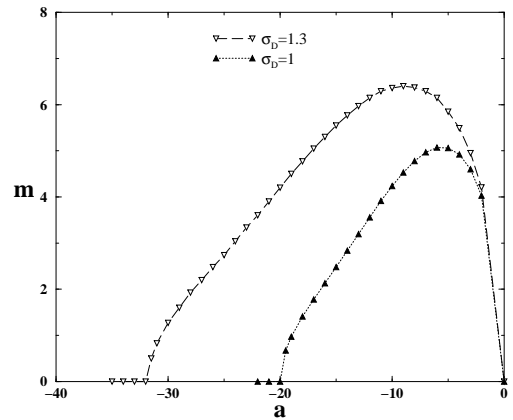,width=8cm}}
\caption{Averaged magnetization of the Langevin equation as a function
of $a$ (i.e. the temperature), in mean field approximation, for different 
values of $\sigma_D$.
}
\end{figure}

 In mean field approximation the order-to-disorder critical point
is located at $a=0$, and in numerical we obtain also a close to zero 
critical value which does not depend on $\sigma_D$.
On the contrary, 
the location of
the disorder-to-order transition, depends on $\sigma_D$, analogously
as the location of $T_1$ depends on $\delta J$ in the AIM. Observe 
that this transition is not sharp in the lowermost curve of Fig.3 
due to finite size effects.
Curves  in Fig.2 and Fig.3 change with increasing $\sigma_D$
in the same way they do in the AIM when increasing $\delta J$, i.e., 
the larger the noise the smaller the ordering.  

\begin{figure}
\centerline{\psfig{figure=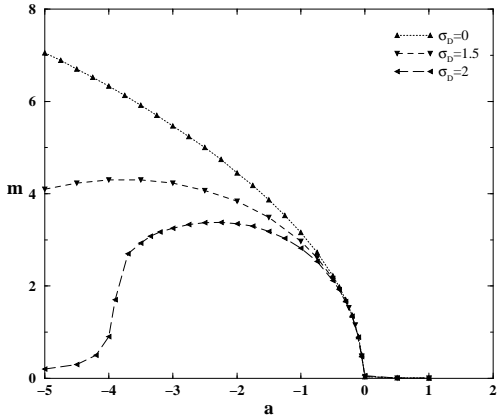,width=8cm}}
\caption{Averaged magnetization of the Langevin equation as a function
of $a$ (i.e. the temperature), in numerical simulations, for different
values of $\sigma_D$.
}
\end{figure}
 Let us stress once more that in order to obtain the transition, we change
both the coefficient of the linear term and of the multiplicative noise
term.
If one of these two coefficients was kept fixed while the other was
changed the microscopic phenomenology would not be reproduced.
{\it The presence of the multiplicative noise term is essential to generate 
the disorder-to-order transition}.
We have performed a numerical study of Eq. (\ref{final})
omitting the term proportional to
$\nabla
 ({\xi}\nabla \psi)$ ,
 and conclude that none of the previous conclusions
is qualitatively affected by this suppression;
by omitting this term the disorder-to-order critical point
is shifted to a lower value of $a$, and consequently this term
has only a disorganizing effect.
We could consequently write down a minimal model just by dropping out 
this unnecessary term, in the same way we omitted other irrelevant
higher order dependences on $D$ in the derivation of the Langevin equation.
We conclude that {\it
  the proposed
Langevin equation with multiplicative noise in the Ito representation
reproduces qualitatively all the interesting
properties of the anneal Ising model, and in particular the
reentrant phase transition.} 
Therefore, once more  
it is shown that the multiplicative noise
 is the key ingredient of highly non-trivial
phenomena in stochastic systems at a mesoscopic level.

  Let us finally remark that the phenomenon we have just described 
{\it is not} the usual noise-induced transition as reported in 
previous works
\cite{Raul,Sancho,GMs}.
First of all, in those works only the multiplicative noise amplitude 
has to be changed to obtain a NIOT, while in our case the transition
is obtained by varying the parameter $a$ that multiplies both
the multiplicative noise and the linear term. Consequently in our case
the disorder-to-order transition is not purely noise-induced.
Second,  
 considering a Stratonovich representation 
of the Langevin equation with multiplicative noise is essential in those
works 
to generate noise-induced ordering.
 In fact, standard Langevin equations as
those described in \cite{Raul,Sancho,GMs} do not exhibit NIOTs 
 when
intended in the Ito representation (\cite{preprint}).
 On the other hand, in the model presented here, 
the Langevin equation is intended
in the Ito sense, and due to its peculiar structure, namely the coupling 
between $a$ and $\xi(t)$, that we have
justified from a microscopic point of view, it 
can exhibit a rather rich
phenomenology.
     In particular the system shows an ordering and a disordering
    transition as the temperature is increased
 but it does not exhibit, for example,
the short time instability characteristic of the phenomena discussed
in \cite{Raul,Kawai,GMs}.  

\vspace{8pt}
{\bf \centerline{ACKNOWLEDGEMENTS}}
\vspace{8pt}

    It is a pleasure to acknowledge 
 L. Pietronero,  M. Scattoni and S. Pellegrini and Juan Ruiz-Lorenzo
 for useful comments and discussions, and J.M.Sancho for a critical reading
of the first draft and interesting correspondence.
This work has been  partially
supported by the European Union through a grant to M.A.M. 
ERBFMBICT960925, and by the TMR 'Fractals' network,  
project number EMRXCT980183.

\end{text}


\begin{references}
\bibitem{preprint} An extended discussion of this point will be presented 
elsewhere.
\end{references}
\end{document}